\RequirePackage[2020-02-02]{latexrelease}
\documentclass[preprintnumbers,amsmath,amssymb,floatfix,10pt,prd,onecolumn,
superscriptaddress,nofootinbib]{revtex4}

\usepackage{amsfonts}
\usepackage{mathrsfs}
\usepackage{latexsym}
\usepackage{epsfig}
\usepackage{epstopdf}
\usepackage{graphicx}
\usepackage{amssymb}
\usepackage{amsmath}
\usepackage{dcolumn}
\usepackage{bm}
\usepackage{color}
\usepackage{xcolor}
\usepackage{comment}
\usepackage[cp1251]{inputenc}
\begin{document}

\title{\bf $f(R, T)$ Gravity Bouncing Universe with Cosmological Parameters}

\author{Adnan Malik}
\email{adnan.malik@zjnu.edu.cn: adnan.malik@skt.umt.edu.pk; adnanmalik_chheena@yahoo.com}
\affiliation{School of Mathematical Sciences, Zhejiang Normal University, Jinhua, Zhejiang, China.}
\affiliation {Department of Mathematics, University of Management and Technology, Sialkot Campus, Pakistan}

\author{Tayyaba Naz}
\email{tayyaba.naz@nu.edu.pk}\affiliation{National University of Computer and
Emerging Sciences,\\ Lahore Campus, Pakistan.}

\author{Aimen Rauf}
\email{aimenrauf90@gmail.com}\affiliation{Department of Mathematics, University of Management and Technology,\\ Sialkot Campus, Lahore, Pakistan}

\author{M Farasat Shamir}
\email{farasat.shamir@nu.edu.pk; mfs24@leicester.ac.uk; farasat.shamir@gmail.com}
\affiliation{National University of Computer and Emerging Sciences,\\ Lahore Campus, Pakistan.}
\affiliation{School of Computing and Mathematical Sciences, \\University of Leicester, United Kingdom}
\author{ Z. Yousaf}
\email{zeeshan.math@pu.edu.pk}\affiliation{Department of Mathematics, University of the Punjab,\\ Quaid-i-Azam Campus, Lahore-54590, Pakistan.}

\begin{abstract}
\begin{center}
\textbf{Abstract}\\
\end{center}

\bigskip
The basic aim of this manuscript is to investigate the cosmological solutions in the context of the modified $f(R, T)$ theory of gravity, where $R$ is the Ricci scalar and $T$ is the trace of the energy-momentum tensor. For our current work, we consider the Friedmann-Robertson-Walker space-time for finding the solutions of field equations. We investigate the nature of universe by considering acceleration expansion of universe, ultra relativistic universe, sub-relativistic universe, dust universe, radiation universe, stiff universe. Moreover, we apply the power law technique by taking two different $f(R, T)$ gravity models to observe the expanding nature of the universe.  The bouncing scenario is also discussed by choosing some particular values of the model parameters and observed the energy conditions, which are satisfied for a successful bouncing model. It is also concluded that some solutions in $f(R, T)$ theory of gravity supports the concept of exotic matter and accelerated expansion of the universe due to a large amount of negative pressure.\\\\


\end{abstract}

\maketitle

\date{\today}

\section{introduction}

The phenomenon of the expansion of the universe stands as a fundamental pillar in modern cosmology, profoundly shaping our understanding of the cosmos at large. The realization that galaxies are receding from one another, forming an intricate cosmic web, has its roots in the ground breaking observations made by astronomers in the early 20th century. The conceptual framework of an expanding universe was solidified through the lens of Albert Einstein's general relativity, which provided the theoretical scaffolding for interpreting the observed redshifts of distant galaxies. Edwin Hubble's empirical verification of the redshift-distance relationship not only validated the notion of cosmic expansion but also laid the groundwork for the formulation of Hubble's Law, a cornerstone in contemporary astrophysics. The implications of an expanding universe extend far beyond the spatial dimensions of galaxies hurtling away from a common origin; it encompasses the very fabric of spacetime itself. The cosmic microwave background radiation, a remnant from the early universe, bears witness to the evolving dynamics of cosmic expansion. As we delve into this intricate cosmic ballet, the introduction of this research article aims to provide a comprehensive backdrop, setting the stage for a nuanced exploration into modified gravity theories and exotic matter as they relate to the underlying mechanisms governing the expansive nature of our cosmos \cite{01}. Linder \cite{1} explored the recent expansion history of the universe promises insights into the cosmological model, the nature of dark energy, and potentially clues to high energy physics theories and gravitation. They also examine the extent to which precision distance-redshift observations can map out the history, including the acceleration-deceleration transition, and the components and equations of state of the energy density. Ellis et al., \cite{2} showed that, in principle, the observed galactic redshifts and microwave background radiation in the universe can be explained by a static spherically symmetric (or SSS) universe model, in which there is a singularity which continually interacts with the universe (rather than the once for all interaction that occurs in the Friedmann-Robertson-Walker, or FRW, universe models). Padmanabhan \cite{3} claimed that it is an accepted practice in cosmology to invoke a scalar field with a potential, when the observed evolution of the universe cannot be reconciled with theoretical prejudices. Wetterich \cite{4} discussed a cosmological model where the universe shrinks rather than expands during the radiation and matter dominated periods and presented a simple model where the masses of particles arise from a scalar ``cosmon” field, similar to the Higgs scalar. Astier \cite{5} reviewed the principle and difficulties of the measurements, the classification and diversity of supernovae, and the physics of explosion and discussed the systematic uncertainties affecting the cosmological conclusions with some emphasis on photometric calibration. Turner \cite{6} argued that fundamental physics beyond the standard model is implicated in both the dark matter and dark energy puzzles: new fundamental particles (e.g., axion or neutralino) and new forms of relativistic energy (e.g., vacuum energy or a light scalar field). Vitenti and Penna-Lima \cite{7} introduced a model-independent method to reconstruct directly the deceleration function via a piecewise function by assuming only an isotropic, homogeneous and flat universe and concluded that distance measurements are currently the most powerful tool to study the expansion history of the universe without specifying its matter content nor any theory of gravitation. Vishwakarma \cite{8} explained the current observations are usually due to an accelerating expansion of the present universe, however, with the present quality of the supernovae Ia data, the allowed parameter space is wide enough to accommodate decelerating models as well. Astier and Pain \cite{9} reviewed the various observational evidences, most of them gathered in the last decade, and the improvements expected from projects currently collecting data or in preparation. Freedman \cite{10} discussed about The Hubble constant, which measures the expansion rate, together with the total energy density of the universe, sets the size of the observable universe, its age, and its radius of curvature.\\

Modified theories of gravity play a crucial role in discussions about the expansion of the universe as they provide alternative frameworks to traditional general relativity. While general relativity has been remarkably successful in describing gravity on cosmological scales, modified theories allow researchers to explore deviations from Einstein's equations. These modifications become particularly relevant when attempting to explain phenomena such as dark energy and the accelerated expansion of the universe. By incorporating additional terms or alternative gravitational theories, researchers can address discrepancies between observational data and theoretical predictions. Investigating modified gravity theories is imperative for a comprehensive understanding of the forces driving cosmic expansion and can potentially unveil novel insights into the nature of dark energy and the cosmic acceleration phenomenon. Some of the modified theories of gravity are such as $f(R)$ \cite{11,12,13,14}, $f(G)$ \cite{15,16,16a}, $f(R,G)$ \cite{17,18}, $f(G,T)$ \cite{20}, $f(Q)$ \cite{21}, $f(R,T)$ \cite{22,23}, $f(R, \phi)$ \cite{24,25}, and $f(R, \phi, X)$ \cite{26,27} modified theories of gravity. Shamir and Malik \cite{28} considered the Friedmann-Robertson-Walker space–time for finding some exact solutions by using different values of equation of state parameter for dust universe, radiation universe, ultra-relativistic universe, sub-relativistic universe, stiff universe, and dark energy universe. The same authors \cite{29} examined some numerical solutions and  bouncing scenario by taking the Klein-Gordon Equation and using distinct values of the equation of state (EoS) parameter and concluded that some solution in $f(R)$ theory of gravity supports the concept of exotic matter and accelerated expansion of the universe due to a large amount of negative pressure. Yousaf along his collaborators \cite{30} worked out the cosmic bounce with a cubic form of scale factor and concluded that the stability of the assumed scale factor model is in evidence of universal expansion and allows the universal bounce by developing the validations of energy conditions. Stachowiak and Szydlowski \cite{31} found the exact solutions for the models with $\rho^2$ correction in terms of elementary functions, and showed all evolutional paths on their phase plane. Caruana \cite{32} studied the types of gravitational Lagrangians which are capable of reconstructing analytical solutions for symmetric, oscillatory, superbounce, matter bounce, and singular bounce settings. Shabani and Ziaie \cite{33} studied classical bouncing solutions in the context of $f (R, T)$ gravity in a flat FLRW background using a perfect fluid as the only matter content and this investigation is based on introducing an effective fluid through defining effective energy density and pressure. Burkmar and Bruni \cite{34} explored the dynamics of Friedmann-Lemaitre-Robertson-Walker (FLRW) cosmologies which consist of dark matter, radiation, and dark energy with a quadratic equation of state. Singh et al., \cite{35} explored the bouncing scenario of a flat homogeneous and isotropic universe by using the reconstruction technique for the power-law parametrization of the Hubble parameter in a modified gravity theory with higher-order curvature and trace of the energy-momentum tensor terms. Zubair and Farooq \cite{36} investigated the bouncing cosmology in a 4D Einstein Gauss-Bonnet gravity and examined various bouncing models such as symmetric bounce, matter bounce, super bounce, and oscillatory bounce. Yousaf et al., \cite{37} studied the bouncing nature of the universe for an isotropic configuration of fluid ${{ \mathcal T }}_{\alpha \beta }$ and Friedmann-Lemaitre-Robertson-Walker metric scheme and concluded that all free parameters are set in a way, to make the supposed Hubble parameter act as the bouncing solution and ensure the viability of energy conditions. Shamir \cite{38} explored the possibility of bouncing solutions by choosing a well-known equation of state parameter and considered a Gauss–Bonnet cosmological model with linear trace term. Odintsov and Paul \cite{39} proposed a new five-parameter entropy function that proves to be singular-free during the entire cosmic evolution of the universe, and at the same time, also generalizes the Tsallis, Barrow, Renyi, Sharma-Mittal, Kaniadakis and Loop Quantum Gravity entropies for suitable limits of the parameters.  Agrawal et al., \cite{41} shown the matter bounce scenario of the universe in an extended symmetric teleparallel gravity and performed the stability analysis with the first order scalar perturbation of the Hubble parameter and matter energy density to verify the stability of the model. Cardoso et al., \cite{42} investigated a simple bouncing geometry, with (outer and inner) apparent horizons but no event horizon and shown that the inner horizon blue shifts radiation, which can lead to a gigantic amplification of energy observable from far away regions. Singh and Bamba
\cite{43} examined the stability of the model and found that all the essential features of bouncing model are satisfied successfully. Liu et al., \cite{43a} investigated the exponential stability problem for a class of Cohen–Grossberg neural networks with Markovian jumping parameter and mixed time-delays. Du along with the collaborators \cite{43b} studied the generalized discrete neutral-type neural network with time-varying delays. Abouelmagd et al., \cite{43c} found the secular solution around the triangular equilibrium points and reduce it to the periodic solution in the frame work of the generalized restricted thee-body problem. \\

Motivated from the above literature, we are interested to discuss the exact solutions and bouncing cosmology in modified $f(R,T)$ theory of gravity. In the paper, the work is organized as: Sec. I provides some basics formulism related to $f(R,T)$ modified theory gravity. Sec. III deals with the cosmological solutions using equation of state (EoS). We discuss the energy conditions and bouncing cosmology in Sec. IV. Finally, we conclude our results in Sec. V.

\section{Basic Formulism of $f(R,T)$ gravity}
The Einstein-Hilbert action in modified $f(R,T)$ gravity \cite{44} is defined as
\begin{equation}\label{1}
S=  \int d^4x \sqrt{-g} \bigg[\frac{1}{16\pi G}f(R,T) +L_m\bigg],
\end{equation}
where $f(R,T)$ is a function of Ricci scalar ($R$) and the trace of energy momentum tensor ($T$), $g$ is determinant of metric tensor $g_{\mu \nu}$ and $L_m$ is Lagrangian matter. By varying the action (\ref{1}) with respect to metric tensor, we get the modified $f (R, T)$ gravity field equation as
\begin{equation}\label{2}
f_{R}(R,T) R_{\mu\nu}-\frac{1}{2} f(R,T)g_{\mu\nu}+(g_{\mu\nu} \Box-\nabla_mu\nabla_\nu)f_R(R,T)= 8\pi T_{\mu\nu}-f_T(R,T)T_{\mu\nu}-f_T (R,T) \Theta_{\mu\nu},
\end{equation}
where $\square$ $\equiv$ $\nabla^\mu$ $\nabla_\mu$ is the D’Alembertian operator, $\nabla_\mu$ symbolizes the covariant derivative, $f=f(R,T)$  $f_R \equiv \frac{\partial f}{\partial R}$ , and $f_T \equiv \frac{\partial f}{\partial T}$. The covariant derivative Eq. (\ref{2}) \cite{45} presents
\begin{equation}\label{3}
\nabla^\mu T_{\mu\nu} = \frac{f_{T}(R,T)}{8\pi-f_{T}(R,T)}\Big[[T_{\mu\nu}+\Theta_{\mu\nu}]\nabla^\mu \ln f_{T}(R,T)+\nabla^\mu \Theta_{\mu\nu}-\frac{1}{2}g_{\mu\nu}\nabla^\mu \mathcal{T}\Big],
\end{equation}
which means that the energy momentum tensor in $f(R,T)$ does not obey the conservation law as in $GR$. Moreover, the energy momentum tensor $T_{\mu \nu}$ for perfect fluid is defined as
\begin{equation}\label{5}
T_{\mu \nu}=(\rho+p) u_\mu u_\nu -pg_{\mu \nu},
\end{equation}
where $\rho$, and $p$ represent the energy density, and pressure respectively. The modified field equations (\ref{2}) can be expressed in a $GR$ form as
\begin{equation}\label{4}
G_{\mu\nu}=\frac{1}{f_{R}(R,T)}\Big[8\pi T_{\mu\nu}+\frac{1}{2}[f(R,T)-R f_{R}(R,T)]g_{\mu\nu}-f_{T}(R,T)(T_{\mu\nu}+\Theta_{\mu\nu})-(g_{\mu\nu}\Box- \nabla_\mu\nabla_\nu)f_{R}(R,T)\big].
\end{equation}
Here, $\Theta_{\mu\nu}$ is known as a scalar expansion which is defined as $\Theta_{\mu\nu}=-2 T_{\mu\nu}-P g_{\mu\nu}$. We take $\mathcal{L}_{m}=-\mathcal{P},$ where $\mathcal{P}=\frac{1}{3}(p_{r}+2 p_{t})$ \cite{44}. For our current work, we consider the Friedmann-Robertson-Walker (FRW) space-time as
\begin{equation}\label{6}
  ds^2=dt^2-a^2(t)[dx^2+dy^2+dz^2],
\end{equation}
where $a(t)$ denotes the scale factor depending upon time. The corresponding Ricci scalar is
\begin{equation}\label{7}
R=-6[2H^2+\dot{H}],
\end{equation}
where dot represents the derivative with respect to time and $H=\frac{\dot{a}}{a}$ is the Hubble parameter. Substituting the energy momentum tensor (\ref{5}) and FRW space-time (\ref{6}) in Eq. (\ref{2}), we get the field equations in term of Hubble parameter as
\begin{equation}\label{8}
\rho=\frac{1}{-2+f_T}[f_{R,00}-2f_{R,00}f_T+f_T(f+6f_R(2H^2+\dot{H}))+2f_R(1+f_T)H^2-(f_{R,0}+4f_{R,0}f_T)H+2f_R(1+2f_T)[H^2+
\dot{H}]],
\end{equation}
\begin{equation}\label{9b}
p=\frac{1}{-2+f_T}[-2f+3f_{R,00}-4f_R(2H^2+\dot{H})+6f_RH^2+9f_{R,0}H+6f_R(H^2+\dot{H})].
\end{equation}
These are complicated and highly non-linear partial differential equations, which can not be solved analytically. So, we solve these equations by imposing some physical assumptions. Furthermore, it would be important to mention here that MATHEMATICA software is used for further analysis.
\section{Cosmological solutions using Equation of State}
In this section, we investigate the exact solutions for FRW model in the $f(R,T)$ modified theory gravity. We also observe the graphical depiction of
energy density and pressure component for two different $f(R,T)$ gravity models by taking power-law technique. We consider acceleration expansion of universe, ultra relativistic universe, sub-relativistic universe, dust universe, radiation universe, stiff universe to discuss the cosmological solutions.

\subsection{Model-I:~$f(R,T)=\gamma(R+T)$}
In this subsection, we discuss the cosmological solutions by taking taking $f(R,T)$ gravity model. Moreover, we will present the detailed graphical analysis for energy density and pressure component for different $EoS$ parameters using acceleration expansion of universe, ultra relativistic universe, radiation universe, sub-relativistic universe, dust universe and stiff universe respectively.
\subsubsection{\textbf{Acceleration Expansion of Universe}}
 We implement $w=-1$ for examining the solution of acceleration expansion of universe. The field Eq. (\ref{8}) and (\ref{9b})  set off as following
\begin{equation}\label{9}
f-2f_{R,00}+6f_R(2H^2+\dot{H})-2\big[f_RH^{2}+2[f_{R,0}H+f_R(H^2+\dot{H})]\big]=0,
\end{equation}
which is non-linear and complicated differential equation involving unknown function, hubble parameter and its derivatives. It may be observed that power law form of scale factor as $a(t)=a_0t^k$, where $a_0$ and $k$ are arbitrary real numbers. After substituting all the values in Eq. (\ref{9}), we finally get the following constraint equation as
\begin{equation}\label{10a}
\frac{2k\gamma\big[8\gamma^{2}+3k(3+7\gamma)\big]}{(1+\gamma)\big[3+\gamma(13+8\gamma)\big]}=0.
\end{equation}
\begin{figure}
\includegraphics[width=5cm,height=5cm]{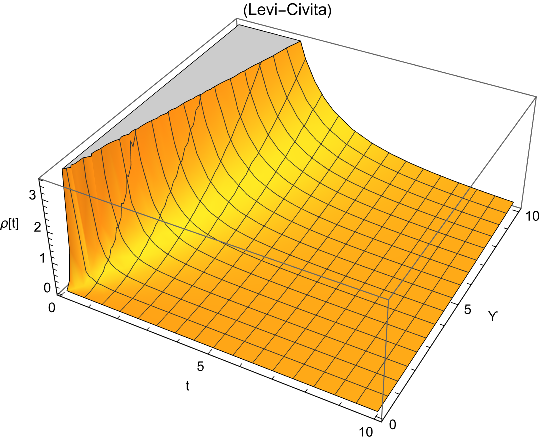}
~~~~~~~~\includegraphics[width=5cm,height=5cm]{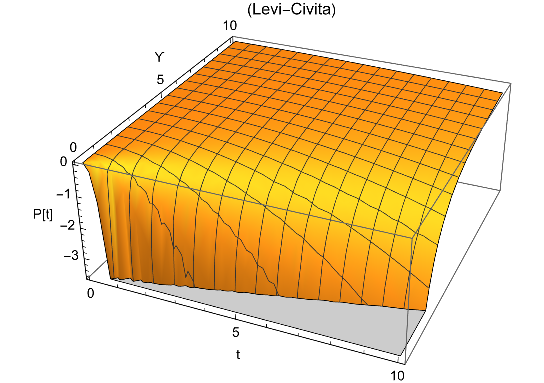}
\caption{\label{Fig.1} Graphical analysis of energy density and pressure component for acceleration expansion of universe using Model-I.}
\end{figure}
We may choose different values of $k$ for finding the solutions of energy density and pressure. If we put $k = 0$ in Eq. (\ref{10a}), then we get the trivial solution. For non-trivial solution, we choose to take $k=\frac{-8\gamma^{2}}{3(3+7\gamma)}$. It can be noticed that the parameter $\gamma$ is very important to discuss the graphical behavior of energy density and pressure component for acceleration expansion of the universe case. By putting different values of the  parameter $\gamma$, we can get different behavior. It can be seen from the Fig. \ref{Fig.1} that the graphical representation of energy density is positive and decreasing while pressure is monotonic i.e. negative as well as positive. The negative nature of pressure indicates the presence of exotic matter, which is the validation of our result that universe is expanding.

\subsubsection{\textbf{Ultra Relativistic Universe}}
\begin{figure}
\includegraphics[width=5cm,height=5cm]{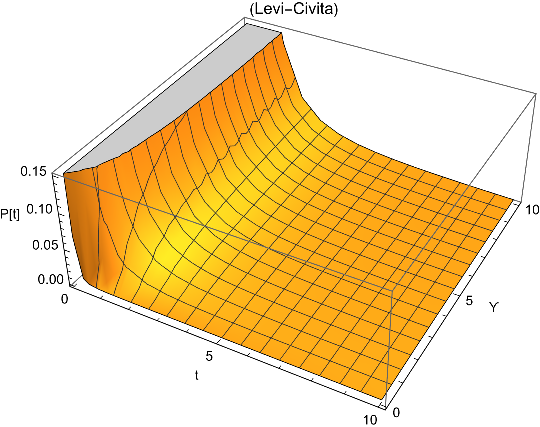}
~~~~~~~~\includegraphics[width=5cm,height=5cm]{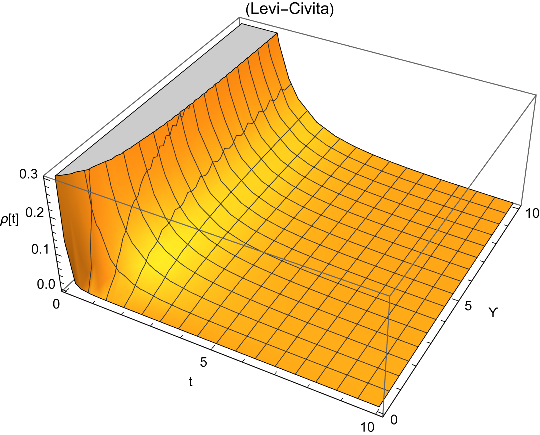}
\caption{\label{Fig.2} Graphical analysis of energy density and pressure component for ultra relativistic universe using Model-I.}
\end{figure}

In Ultra relativistic universe, we use $w=\frac{1}{2}$ and then subtracting the field equations (\ref{8}) and (\ref{9b}) gives
\begin{equation}\label{11}
\begin{split}
\frac{1}{2(-2+f_T)}\Big[-\big[f(4+f_T)-f_{R,00}(5+2f_T)+6f_R(4+f_T)(2H^2+\dot{H})+\\2f_R(7+f_T)H^2+[f_{R,0}(19+5f_T)H+2f_R(5+2f_T
)(H^2p+\dot{H})]\big]\Big]=0,
\end{split}
\end{equation}
Substituting the power law and $f(R,T)$ gravity model in Eq. (\ref{11}), we get
\begin{equation}\label{12}
\frac{k\gamma[2\gamma(3+5\gamma)-3k[3+\gamma(7+6\gamma)]]}{(1+\gamma)[3+\gamma(13+8\gamma)]}=0.
\end{equation}
We consider $k=\frac{2(3\gamma+5\gamma^{2})}{3(3+7\gamma+6\gamma^{2})}$ for obtaining the constraint equation. The graphical analysis of energy density and pressure component are similar in nature i.e., both components are positive and decreasing as shown in Fig. \ref{Fig.2}.

\subsubsection{\textbf{Sub-Relativistic Universe}}
For sub-relativistic universe, we use $w=\frac{1}{4}$ and then subtracting the field equations (\ref{8}) and (\ref{9b}) gives
\begin{equation}\label{15}
\begin{split}
\frac{1}{4(-2+f_T)}[-(f(8+f_T)-f_{R,00}(11+2f_T)+6f_R(8+f_T)(2H^2+\dot{H}))+\\2f_R(13+f_T)H^2
+f_{R,0}(37+4f_T)H+2f_R(11+2f_T)(H^2+\dot{H})]=0.
\end{split}
\end{equation}
Substituting the values in Eq. (\ref{15}), we get
\begin{equation}\label{16}
\frac{k\gamma[2\gamma(5+11\gamma)-3k(3+\gamma(7+10\gamma))]}{4(1+\gamma)[3+\gamma(13+8\gamma)]}=0.
\end{equation}
For the sake of simplicity, we choose $k=\frac{2(5\gamma+11\gamma^{2})}{3(3+7\gamma+10\gamma^{2})}$ for discussing the solutions in Sub-Relativistic universe. The graphical behavior of sub-relativistic universe is exactly same as ultra relativistic universe. That's why we did not show the graphical depiction for this case.

\subsubsection{\textbf{Dust Universe}}
\begin{figure}
\includegraphics[width=5cm,height=5cm]{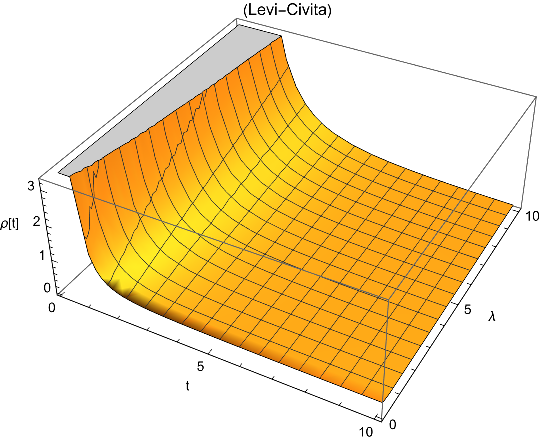}
~~~~~~~~\includegraphics[width=5cm,height=5cm]{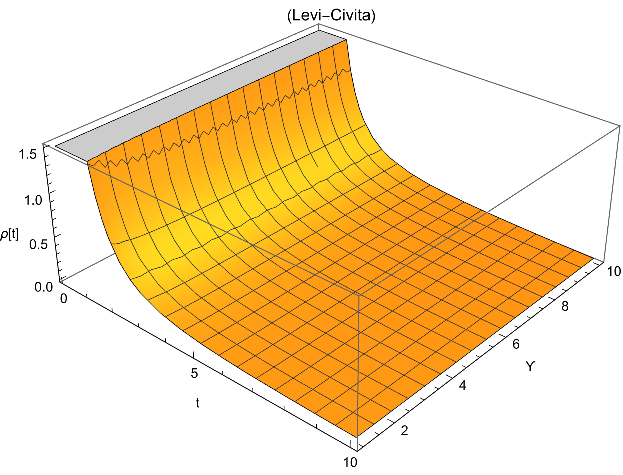}
\caption{\label{Fig.3} Graphical analysis of energy density and pressure component for dust universe using Model-I}
\end{figure}

For this case, we use $w=0$ and then solve the field equations (\ref{8}) and (\ref{9b}), we get
\begin{equation}\label{17}
\frac{[-2f+3[f_{R,00}-4f_R(2H^2+\dot{H})]]+6f_RH^2+[9f_{R,0}H+6f_R(H^2+\dot{H})]}{-2+f_T}=0,
\end{equation}
and the constraint equation for this case is as follows
\begin{equation}\label{18}
\frac{4k\gamma^{2}(1-3(-1+k)\gamma)}{(1+\gamma)[3+\gamma(13+8\gamma)]}=0.
\end{equation}
For satisfying the constraint equation, we choose $k=\frac{1+3\gamma}{3\gamma}$ with $\gamma \neq 0$, which give us as a good results of density and pressure. The graphical behavior of energy density and pressure component show decreasing and positive behavior as shown in Fig. \ref{Fig.3}.

\subsubsection{\textbf{Radiation Universe}}
\begin{figure}
\includegraphics[width=5cm,height=5cm]{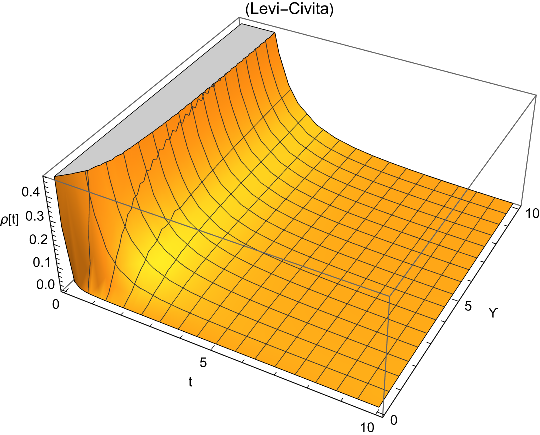}
~~~~~~~~\includegraphics[width=5cm,height=5cm]{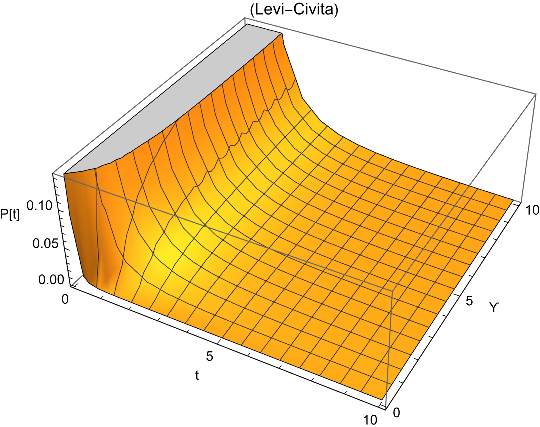}
\caption{\label{Fig.4} Graphical analysis of energy density and pressure component for radiation universe using Model-I}
\end{figure}

For radiation universe, we apply $w=\frac{1}{3}$ to the field equations (\ref{8}) and (\ref{9b}) and get
\begin{equation}\label{13}
\begin{split}
\frac{1}{3(-2+f_T)}[2f_{R,00}(4+f_T)-f(6+f_T)-6f_R(6+f_T)(2H^2+\dot{H})]\\+2f_R(10+f_T)H^2+[f_{R,0}(7+f_T)H+f_R(4+f_T)
(H^2+\dot{H})]=0,
\end{split}
\end{equation}
and the corresponding constraint equation is
\begin{equation}\label{14}
\frac{-2k\gamma\big[-8\gamma(1+2\gamma)+3k[3+\gamma(7+8\gamma)]\big]}{3(1+\gamma)\big[3+\gamma(13+8\gamma)\big]}=0.
\end{equation}
By selecting different values of $k$, we can get the interesting results of the constraint equation. For the sake of our current analysis, we choose $k=\frac{8(\gamma+2\gamma^{2})}{3(3+7\gamma+8\gamma^{2})}$ and investigate the behavior of pressure and energy density, which is similar to $\omega=\frac{1}{2}$. The graphical analysis of energy density and pressure component is positive and decreasing as shown in Fig. \ref{Fig.4}.

\subsubsection{\textbf{Stiff universe}}
For this case, we use $w=1$ and solving the field equations (\ref{8}) and (\ref{9b}) as
\begin{equation}\label{19}
\begin{split}
\frac{1}{-2+f_T}[-(-2f_{R,00}(1+f_T)+f(2+f_T)+6f_R(2+f_T)(2H^2+\dot{H}))+\\2f_R(4+f_T)H^2
+2(f_{R,0}(5+2f_T)H+2f_R(1+f_T)(H^2+\dot{H}))]
=0,
\end{split}
\end{equation}
and the constraint equation is as following
\begin{equation}\label{20}
\frac{-2k\gamma(-4\gamma+3k(3+4\gamma))}{(3+\gamma(13+8\gamma))}=0.
\end{equation}
For stiff universe, we choose $k=\frac{4\gamma}{3(3+4\gamma)}$ to satisfies the constraint equation. The graphical analysis of energy density and pressure component are similar in nature i.e., both components are positive and decreasing. Moreover, their behavior is exactly same like ultra-relativistic universe, so we did not show the plots due to repetition.

\subsection{Model-II:~$f(R,T)=R+\alpha\lambda[(1+\frac{R^{2}}{\lambda^{2}})^{-n}-1]+\gamma T$}
In this subsection, we discuss the cosmological solutions by taking taking $f(R,T)=R+\alpha\lambda[(1+\frac{R^{2}}{\lambda^{2}})^{-n}-1]+\gamma T$ gravity model. Moreover, we will present the detailed graphical analysis for energy density and pressure component for different $EoS$ parameters using acceleration expansion of universe, ultra relativistic universe, radiation universe, sub-relativistic universe and Stiff universe respectively. We also use power law technique to determining the behavior of the universe.

\subsubsection{\textbf{Acceleration Expansion of Universe}}
\begin{figure}
\includegraphics[width=5cm,height=5cm]{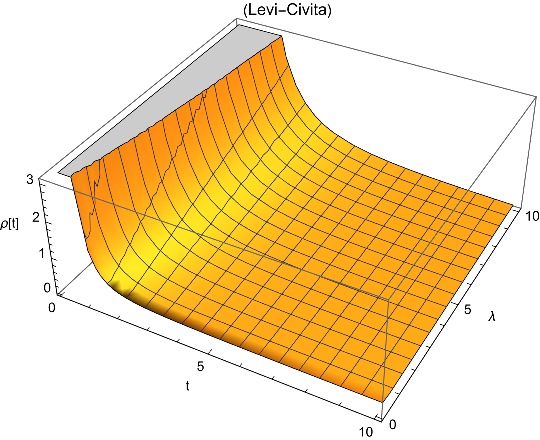}
~~~~~~~~\includegraphics[width=5cm,height=5cm]{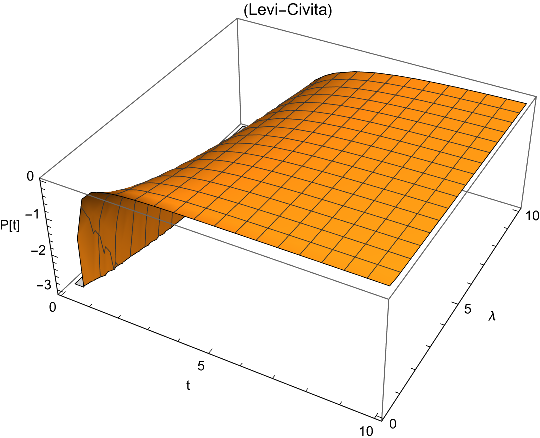}
\caption{\label{Fig.5} Graphical analysis of energy density and pressure component for acceleration expansion of universe using Model-II}
\end{figure}

For the purpose of examining the acceleration expansion of universe, we use $w= -1$ and field equations (\ref{8}) and (\ref{9b}) become
\begin{equation}\label{22}
\begin{split}
\frac{1}{(1+\gamma)[3+\gamma(13+8\gamma)]}[-3R+3\alpha\lambda-7R\gamma+7\alpha\lambda\gamma+8f_{R,00}\gamma^{2}
-\alpha(1+\frac{R^{2}}{\lambda^{2}})^{-n}\lambda(3+7\gamma)\\+f_{R}(3R+7R\gamma-16\dot{H}\gamma^{2}+6H^{2}(3+7\gamma))
+2f_{R,0}H[9+\gamma(21+(12-8H)\gamma)]]=0.
\end{split}
\end{equation}
The above differential equation (\ref{22}) is a very complicated and non-linear differential equation involving some unknown function, hubble parameter and derivatives. After substituting all the values in Eq. (\ref{22}), we can get non-linear differential equation, which can be seen in Appendix A. For determining the solutions of energy density and pressure, we have different choices of parameters. But for simplicity, we choose $\gamma=\frac{3}{32}(21\lambda+\sqrt{3}\sqrt{\lambda(-64+147\lambda)})$ and $\alpha=\frac{4}{5}$ and get the following equation as
\begin{equation}\label{100}
9k(2k+3\lambda)\big[32+14\lambda+7\sqrt{3}\sqrt{\lambda(-64+147\lambda)}\big]
=0.
\end{equation}
The graphical analysis of energy density is positive and decreasing while the graphical representation of pressure is negative and increasing as shown in Fig. \ref{Fig.5}. The negative nature of pressure indicates the presence of exotic matter, which is the validation of our result that universe is expanding.

\subsubsection{\textbf{Ultra relativistic universe}}
\begin{figure}
\includegraphics[width=5cm,height=5cm]{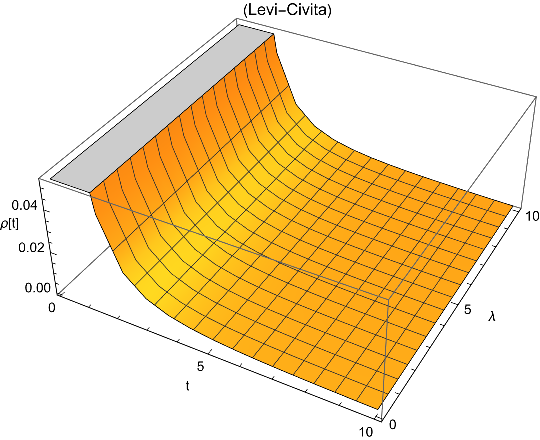}
~~~~~~~~\includegraphics[width=5cm,height=5cm]{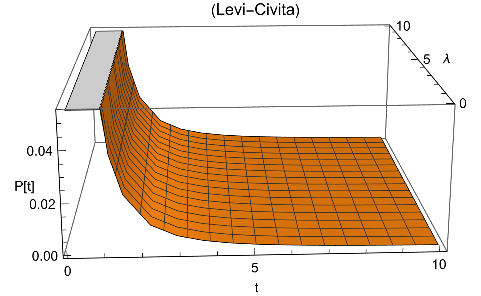}
\caption{\label{Fig.6}  Graphical analysis of energy density and pressure component for ultra relativistic universe using Model-II}
\end{figure}

For ultra-relativistic universe, we utilize $w=\frac{1}{2}$ in field equations (\ref{8}) and (\ref{9b}) and get
\begin{equation}\label{11a}
\begin{split}
\frac{1}{2(1+\gamma)[3+\gamma(13+8\gamma)]}\big[-3R+3f_{R}(6H^{2}+R)+3\alpha\lambda+(-6f_{R,00}+42f_{R}H^{2}+12f_{R}\dot{H}-7R+7f_{R}R+
\\7\alpha\lambda)\gamma+2(-5f_{R,00}+18f_{R}H^{2}+10f_{R}\dot{H}-3R+3f_{R}R+3\lambda)\gamma^{2}+2f_{R,0}H(9+6(2+H)\gamma+\\(3+10H)\gamma^{2})
-\alpha(1+\frac{R^{2}}{\lambda^{2}})^{-n} \lambda[3+\gamma(7+6\gamma)]\big]=0
\end{split}
\end{equation}
Substituting all the value in the above differential equation (\ref{11a}) and get another differential equation as expressed in Appendix A2.
For the discussion of the cosmological solutions for this case, we choose $n=\lambda$, $\alpha=\frac{-9}{10}$, $k=\frac{1}{2}$ and $\gamma=\frac{3}{4}(3+\sqrt{17})$ respectively. The graphical analysis of energy density and pressure component is positive and decreasing for ultra relativistic universe, which can be seen in Fig. \ref{Fig.6}.
\subsubsection{\textbf{Radiation Universe}}
\begin{figure}
\includegraphics[width=5cm,height=5cm]{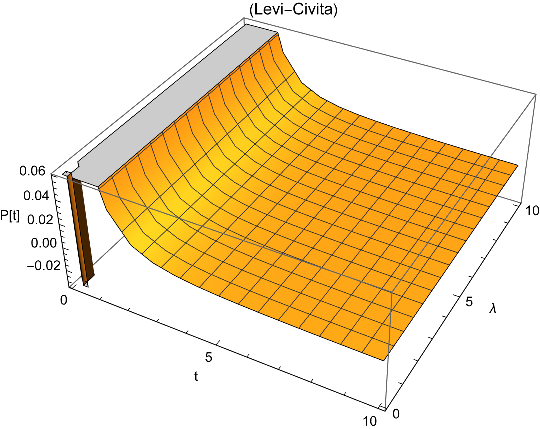}
~~~~~~~~\includegraphics[width=5cm,height=5cm]{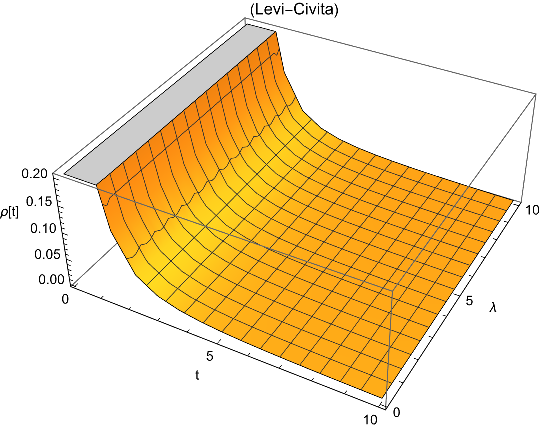}
\caption{\label{Fig.7} Graphical analysis of energy density and pressure component for radiation universe using Model-II}
\end{figure}

Considering $w=\frac{1}{3}$ for the radiation universe and applying the field equation yields the following equation
\begin{equation}\label{27}
\begin{split}
\frac{1}{3(1+\gamma)[3+\gamma(13+8\gamma)]}\Big(-3[-R+f_{R}(6H^{2}+R)+\alpha\lambda]+(8f_{R,00}-42f_{R}H^{2}-16f_{R}\dot{H}+7R\\-7f_{R}R-
7\alpha\lambda)\gamma-8[-2f_{R,00}-R+f_{R}(6H^{2}+4\dot{H}+R)+\alpha\lambda]\gamma^{2}+\alpha(1+\frac{R^{2}}{\lambda^{2}})^{-n}
\lambda[3+\gamma(7+8\gamma)]\\-2f_{R,0}H[9+\gamma[9+8H(1+2\gamma)]]\Big)=0.
\end{split}
\end{equation}
Putting all the values into Eq. (\ref{27}), we get Eq. (\ref{16a}), which is presented in Appendix. For our current analysis, we use $n=-1$, $\alpha=\frac{4}{5}$, $k=\frac{1}{2}$ and $\gamma=\frac{5+\sqrt{313}}{16}$. The graphical analysis of energy density is positive while pressure component is negative near the origin and becomes positive when we move away from the center as seen in Fig. \ref{Fig.7}.

\subsubsection{\textbf{Sub-Relativistic Universe}}
We choose case $w=\frac{1}{4}$ for the sub-relativistic universe and substitute the field equation to obtain the following equation
\begin{equation}\label{29}
\begin{split}
\frac{1}{4(1+\gamma)[3+\gamma(13+8\gamma)]}\Big(-3[-R+f_{R}(6H^{2}+R)+\alpha\lambda]+(10f_{R,00}-42f_{R}H^{2}-20f_{R}\dot{H}+7R\\-7f_{R}R-
7\alpha\lambda)\gamma-2(-11f_{R,00}+30f_{R}H^{2}+22f_{R}\dot{H}-5R+5f_{R}R+5\alpha\lambda)\gamma^{2}+\alpha(1+\frac{R^{2}}{\lambda^{2}})^{-n}\lambda
[3\\+\gamma(7+10\gamma)]-2f_{R}H[9+\gamma[6-3\gamma+2H(5+11\gamma)]]\Big)=0,
\end{split}
\end{equation}
Putting all the values in Eq. (\ref{29}), we get a differential equation (\ref{30a}) in Appendix. We consider $k=\frac{1}{2}$, $\alpha=9$, $n=-1$ and $\gamma=\frac{(1+\sqrt{505})}{28}$ for investigating sub-relativistic universe. For this case, if we select $\gamma$ as the negative then energy density shows negative behavior. To avoid the negativity of energy density, we choose the positive values of $\gamma$. The nature of energy density and pressure component is exactly same as radiation universe.

\subsubsection{\textbf{Stiff universe}}
In order to investigate the nature of stiff universe, the field equations (\ref{8}) and (\ref{9b}) takes the form as
\begin{equation}\label{19}
\begin{split}
\frac{-1}{3+\gamma(13+8\gamma)}[-3R+3f_{R}(6H^{2}+R)+3\alpha\lambda+4f_{R}(6H^{2}+\dot{H}+R)\gamma-4(f_{R,00}\\+R-\alpha\lambda)\gamma-\alpha(1+\frac{R^{2}}{\lambda^{2}})^{-n}\lambda
(3+4\gamma)+[f_{R,0}H(6+4H)\gamma]]=0.
\end{split}
\end{equation}
After putting the values in Eq. (\ref{19}), we get the expression (\ref{31a}) as written in Appendix A. for this case, We choose $\gamma=\frac{-3\lambda}{28(-2+3\lambda)}$, and $\alpha=9$ and the corresponding constraint equation is as follows:
\begin{equation}\label{20}
\frac{2k(-56+81\lambda)\big[\lambda+6k(-7+10\lambda)\big]}{7\lambda(-2+3\lambda)}=0.
\end{equation}
Moreover, we select $k=\frac{-\lambda}{6(-7+10\lambda)}$ for discussing the nature of stiff universe. The graphical depiction of energy density and pressure component is positive as shown in Fig. \ref{8}.
\begin{figure}
\includegraphics[width=5cm,height=5cm]{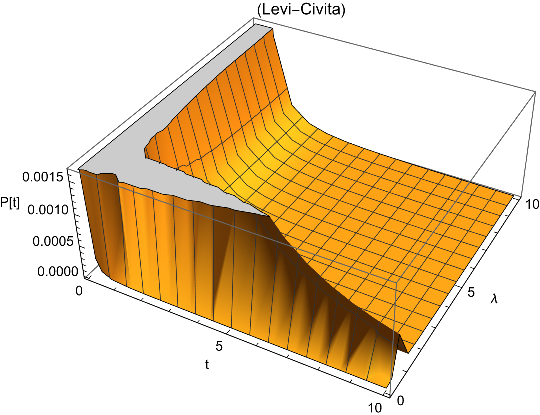}
~~~~~~~~\includegraphics[width=5cm,height=5cm]{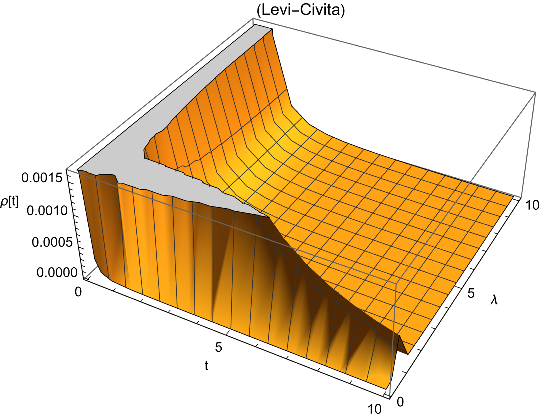}
\caption{\label{Fig.8} Graphical analysis of energy density and pressure component for stiff universe using Model-II}
\end{figure}

\section{Bouncing cosmology in $f(\mathcal{R},\mathcal{T})$ gravity}
\begin{figure}
\includegraphics[width=4.5cm,height=4.5cm]{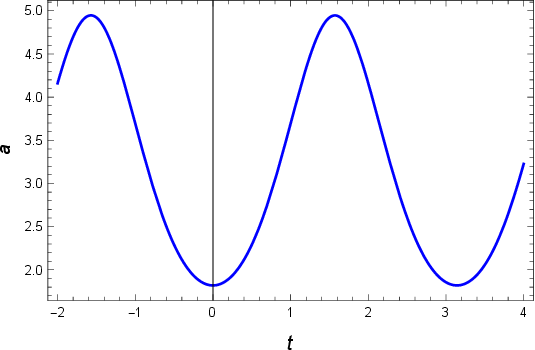}
~~~~~~~~\includegraphics[width=4.5cm,height=4.5cm]{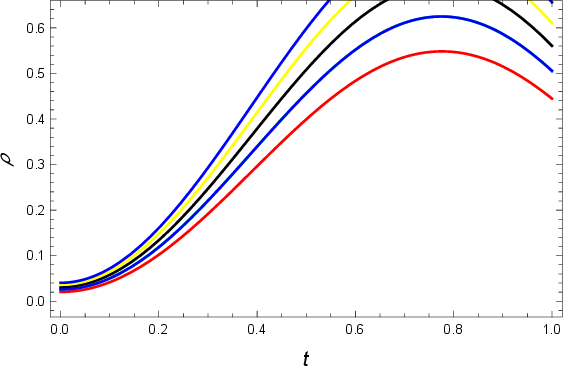}
~~~~~~\includegraphics[width=4.5cm,height=4.5cm]{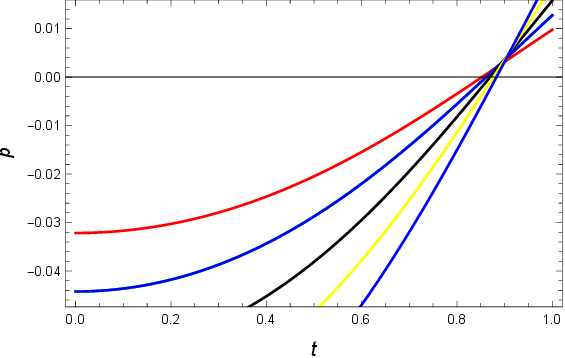}
\caption{\label{Fig.9} Evolution of $a(t)$, energy density and pressure component for bouncing cosmology}
\end{figure}

\begin{figure}
\includegraphics[width=4.5cm,height=4.5cm]{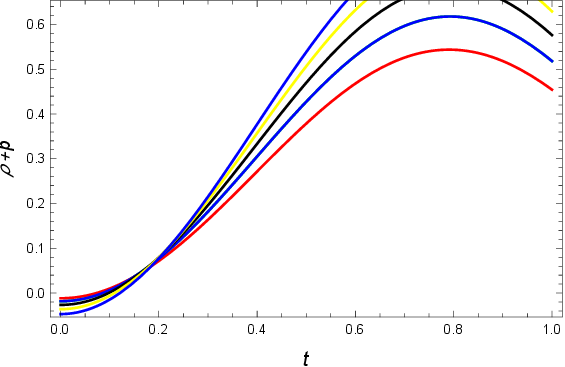}
~~~~~~\includegraphics[width=4.5cm,height=4.5cm]{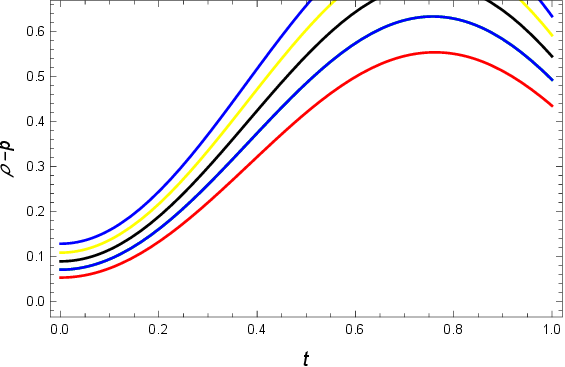}
\caption{\label{Fig.10} Behavior of energy conditions for bouncing cosmology}
\end{figure}

The concept of a bouncing universe in the background of modified theories of gravity represents a captivating departure from the traditional narrative of cosmic origins. Unlike the conventional big bang singularity, where the universe emerges from an infinitesimally small and dense state, the bouncing universe scenario envisions a cyclical cosmic evolution. In these modified gravity theories, the universe undergoes a contraction phase, during which the scale factor diminishes to a finite volume, followed by an expansion phase akin to a cosmic rebound. This bouncing phenomenon is proposed as a remedy to the singularity problem associated with the big bang, providing a more complete and non-singular description of the universe's lifecycle. The crucial element in facilitating a successful bounce is the violation of the null energy condition (NEC) around the bouncing point. By relaxing the energy conditions, these theories allow for exotic forms of matter or energy that drive the contraction and subsequent expansion, offering an intriguing alternative framework for understanding the cosmos. While still theoretical and subject to ongoing research, the concept of a bouncing universe reflects the curiosity and innovation within cosmology as scientists explore novel avenues to refine our understanding of the universe's origins and evolution. Energy conditions are defined as

\begin{eqnarray*}
   NEC &\Rightarrow& \rho+p \geq 0 ,\\
   WEC &\Rightarrow& \rho \geq 0,  \rho+p \geq 0, \\
   SEC &\Rightarrow&  \rho +p \geq 0, \rho+3p \geq 0, \\
   DEC &\Rightarrow&  \rho \geq 0,  \rho -p \geq 0,\\
\end{eqnarray*}

where $NEC$, $WEC$, $SEC$ and $DEC$ denote the null energy conditions, weak energy conditions, strong energy conditions and dominating energy conditions respectively. Moreover, the scale factor $a(t)$ is considered as
\begin{equation}\label{34}
a(t)=\kappa Exp[\alpha\frac{e^{\zeta t}[\zeta sin(\xi t)-\xi cos(\xi t)]}{\zeta^2+\xi^2}].
\end{equation}
where $\kappa$ is an arbitrary integration constant. Now, using Eq. (\ref{34}) in field equations (\ref{8}) and (\ref{9b}), we get
\begin{equation}\label{35}
\rho=\frac{2e^{t\zeta}\alpha\gamma[-2\xi(-1+\gamma)\gamma \cos[t\xi]+\sin[t\xi]\big[-2\zeta(-1+\gamma)\gamma+3e^{t\zeta}\alpha(3+7\gamma+2\gamma^2)\sin[t\xi]\big]]}{(1+\gamma)(3+13\gamma+8\gamma^2)},
\end{equation}
\begin{equation}\label{36}
p=-\frac{4e^{t\zeta}\alpha\gamma^2[\xi(1+3\gamma)\cos[t\xi]+\sin[t\xi](\xi+3\xi\gamma+3e^{t\zeta}\alpha\gamma \sin[t\xi])]}{(1+\gamma)(3+13\gamma+8\gamma^2)}.
\end{equation}
The graphical representation of the scale factor shows that there exist two bouncing points. Scale factor has the minimum values at the beginning bouncing point at
$t = 0$. Accordingly, it can be claimed that our solution comes up with a picture of the evolution of the universe.  For $t < 0$, it shows contracting phase, bounce at $t = 0$
and shows the expanding phase for $t< 0$.  Similar behavior can be observed near the other bounce point as shown in Fig. \ref{Fig.9}. The graphical behavior of energy density is zero at the origin and then shows the increasing behavior when moves towards the boundary as shown in the middle panel of Fig. \ref{Fig.9}. The right panel of Fig. \ref{Fig.9} represents the graphical nature of pressure component, which is initially negative and becomes positive with the movement towards the boundary. It can be claimed that negative nature of pressure component may support the presence of exotic matter, which is a crucial candidate for expanding the universe. The graphical analysis of $\rho+p$ is initially negative near the center and then becomes positive as well as increasing, which can be seen in the left portion of Fig. \ref{Fig.10}, which means that NEC is violated. It can be noticed that WEC and SEC are directly linked with NEC, which means that WEC and SEC will be violated with the violation of NEC. Hence, one can say concluded that all three energy conditions like NEC, WEC and SEC are not satisfied for this case. Furthermore, the difference of energy density and pressure component is positive, which means that DEC is satisfied. It can be concluded that negative nature of pressure component and the violation of energy conditions specially NEC may support the presence of exotic matter, which is a crucial candidate for expanding the universe.

\section{Conclusion}
In this work, the cosmological solutions in the background of the modified $f(R, T)$ theory of gravity by considering FRW space-time for perfect fluid are investigated. For this work, the nature of universe by considering acceleration expansion of universe, ultra relativistic universe, sub-relativistic universe, dust universe, radiation universe and stiff universe are examined. Moreover, two different $f(R, T)$ gravity models by applying the power law technique are chosen for the investigation of solutions. A detailed graphical analysis is provided for discussing the obtained bouncing solutions. The main results of present study are itemized as follows.

In order to discuss the cosmological solutions for first $f(R, T)$ gravity model, we will present the detailed discussions of the nature of energy density and pressure component. The graphical representation of energy density is positive and decreasing while pressure is monotonic i.e. negative as well as positive for acceleration expansion of universe. For, ultra relativistic universe, the graphical analysis of energy density and pressure component are similar in nature i.e., both components are positive and decreasing. The graphical depiction of energy density and pressure component is positive and decreasing for radiation universe. It is interesting to mention here that the graphical behavior of energy density density and pressure component of sub-relativistic universe and stiff universe are exactly same as ultra relativistic universe. It can be concluded that the negative nature of pressure indicates the presence of exotic matter, which is the validation of our result that universe is expanding.

For the discussion of second $f(R, T)$ gravity model, we will explain the detailed discussions related to energy density and pressure component. The graphical analysis of energy density is positive and decreasing while the graphical representation of pressure is negative and increasing for acceleration expansion of universe. The negative nature of pressure indicates the presence of exotic matter, which is the validation of our result that universe is expanding. The graphical analysis of energy density and pressure component is positive and decreasing for ultra relativistic universe. For radiation universe, the graphical analysis of energy density is positive while pressure component is negative near the origin and becomes positive when we move away from the center. It is interesting to mention here that the nature of energy density and pressure component for sub-relativistic universe is exactly same as radiation universe. For stiff universe, the graphical analysis of energy density and pressure component are positive.

To examine the bouncing cosmology, we imposed few restrictions on the parameters, which are used in the model. Our model shows two bounce: the first bounce is at
$t = 0$ and the other is on $t = 3.2$ by taking these limitations. One can get some different bouncing behavior by choosing different values of parameters. It can also be noticed that energy density has positive and increasing behavior, but the pressure component has negative trends near the origin. It can be claimed that NEC, WEC and SEC are violated due to negative behavior of $\rho+p$ near the center. Moreover, the graphical representation of $\rho-p$ is positive, which means that DEC is satisfied. It can be concluded that negative nature of pressure component and the violation of energy conditions specially NEC may support the presence of exotic matter, which is a crucial candidate for expanding the universe.

All the above itemized discussions suggest that the proposed $f(R, T)$ gravity model provides good bouncing solutions with the chosen EoS parameters.

\appendix
\section{}
\begin{equation}\label{23A}
\begin{split}
\frac{1}{(1+\gamma)[3+\gamma(13+8\gamma)]}\Big[3\alpha\lambda+7\alpha\lambda\gamma+\frac{18(\dot{a}^2+a\ddot{a})}{a^{2}}+\frac{42\gamma(\dot{a}^2+a\ddot{a})}{a^{2}}
-\alpha\lambda(3+7\gamma)(1+\\\frac{36(\dot{a}^2+a\ddot{a})^{2}}{\lambda^{2}a^{4}})^{-n}+\frac{1}{\lambda a^{4}} 2(8\gamma^{2}\dot{a}^{2}-(9+\gamma(21+8\gamma))a\ddot{a})\Big[\lambda a^{2}
+12n\alpha(\dot{a}^2+a\ddot{a})(1+\frac{36(\dot{a}^2+a\ddot{a})^{2}}{\lambda^{2} a^{4}})-
(24n\alpha\lambda\dot{a}(3\\(1+\gamma)(3+4\gamma)a-8\gamma^{2}\dot{a})(-\lambda^{2} a^{4}+36(1+2n)\dot{a}^{4}+
72(1+2n)a\dot{a}^{2}\ddot{a}+36(1+2n)a^{2}\ddot{a}^{2})(1+\frac{36(\dot{a}^2+a\ddot{a})^{2}}{\lambda^{2}
a^{4}})^{-n}(-2\\\dot{a}^{3}+a\dot{a}\ddot{a}+a^{2}\dddot{a}))/(a(\lambda^{2}a^{4}+36\dot{a}^{4}+72a\dot{a}^{2}\ddot{a}+36a^{2}\ddot{a}^{2})^{2})
+\frac{1}{\lambda^{5}\dot{a}^{2}}96na\gamma^{2}(1+\frac{36(\ddot{a}+a\ddot{a})^{2}}{\lambda^2a^{4}})^{-3-n}\Big(6\dot{a}^{2}(\dot{a}^{2}+a\ddot{a})\\
(\lambda^{2}a^{4}+36(\dot{a}^{2}+a\ddot{a}^{2})^{2})-2a\ddot{a}(\dot{a}^{2}+a\ddot{a})(\lambda^{2}a^{4}+36(\dot{a}^{2}+a\ddot{a}^{2})^{2})
-4a\dot{a}(\lambda^{2}a^{4}+36(\dot{a}^{2}+a\ddot{a}^{2})^{2})
(3\dot{a}\ddot{a}+aa^{3})-288(\\1+n)\dot{a}(\dot{a}^{2}+a\ddot{a})^{2}(\lambda^{2}a^{4}+36(\dot{a}^{2}+a\ddot{a}^{2}))(2\dot{a}^{3}
+a\dot{a}\ddot{a}+a^{2}\dddot{a})^{2}+144(1+n)a(\dot{a}^{2}+a\ddot{a})(\lambda^{2}a^{4}+36(\dot{a}^{2}+a\ddot{a})^{2})(3\dot{a}\ddot{a}+\\a\dddot{a})(2\dot{a}^{3}
+a\dot{a}\ddot{a}+a^{2}\dddot{a})^{2}+5184(1+n)(2+n)(\dot{a}^{2}+a\ddot{a})^{3}(-2\dot{a}^{3}+a\dot{a}\ddot{a}+a^{2}\dddot{a})^{2}
+a^{2}[\lambda^{2}a^{4}+36(\dot{a}^{2}+a\ddot{a}^{2})^{2}(3\ddot{a}^{2}\\+4\dot{a}\ddot{a}+a\dddot{a})-72(1+n)(\dot{a}^{2}+a\ddot{a})(\lambda^{2}a^{4}
+36\dot{a}^{4}+72\dot{a}^{2}\ddot{a}+36a^{2}\ddot{a}^{2})[10\dot{a}^{2}-6a\dot{a}^{4}\ddot{a}-4a^{2}\dot{a}^{3}\ddot{a}+2a^{3}\dot{a}\ddot{a}\dddot{a}
+a^{2}\dot{a}^{2}
(-6\\\dot{a}^{2}+a\dddot{a})+a^{3}(\ddot{a}+a\dddot{a}^{2}+a\ddot{a}\dddot{a})]]\Big)\Big]\Big]=0,
\end{split}
\end{equation}
\\
\\
\begin{equation}\label{12a}
\begin{split}
\frac{1}{2\lambda^{5}(1+\gamma)[3+\gamma(13+8\gamma)]\dot{a}^{2}}\gamma((-1+\gamma)(\frac{1}{(-1+\gamma)\gamma}\lambda^{4}a^{8}
(2(2(-1+\gamma)\gamma\dot{a}^{2}-(9+\gamma(19+8\gamma))a\ddot{a})(\lambda a^{2}+\\12n\alpha a^{2}(6(\dot{a}^{2}+a\ddot{a})+\alpha\lambda a^{2}(1-(1+\frac{36(\dot{a}^{2}+a\ddot{a})^{2}}{\lambda^{2}a^{4}})^{-n}))-(24\alpha\lambda^{2}a^{3}\dot{a}(9(1+\gamma)^{2}a-2(-1+
\gamma)\gamma\dot{a})(-\lambda^{2}a^{4}\\+36(1+2n)\dot{a}^{4}+72(1+2n)a\dot{a}^{2}\ddot{a}+36(1+2n)a^{2}\ddot{a}^{2})(1+\frac{36(\dot{a}^{2}+a\ddot{a})^{2}}{\lambda^{2}a^{4}})^{-n}
(-2\dot{a}^{3}+a\dot{a}\ddot{a}+a^{2}\dddot{a}))/(\lambda^{2}a^{4}+\\36\dot{a}^{4}+72a\dot{a}\ddot{a}+36a^{2}\ddot{a}^{2})^{2})+24n
\alpha(1+\frac{36(\dot{a}^{2}+a\ddot{a})^{2}}{\lambda^{2}a^{4}})^{-3-n}(6\dot{a}(\dot{a}^{2}+a\ddot{a})^{2})^{2}-2a\ddot{a}(\dot{a}^{2}+a\ddot{a})
(\lambda^{2}a^{4}+36(\dot{a}^{2}\\+a\ddot{a})^{2})^{2}(3\dot{a}\ddot{a}+a\dddot{a})-288(1+n)\dot{a}(\dot{a}^{2}+a\ddot{a})^{2}(\lambda^{2}a^{4}+36(\dot{a}^{2}+a\ddot{a})^{2})
(2\dot{a}^{3}-a\dot{a}\ddot{a}-a^{2}\dddot{a})+144(1+n)a(\dot{a}^{2}\\+a\ddot{a})(\lambda^{2}a^{4}+36(\dot{a}^{2}+a\ddot{a})^{2})(3\dot{a}\ddot{a}+
a\dddot{a})(2\dot{a}^{3}-a\dot{a}\ddot{a}-a^{2}\dddot{a})+5184(1+n)(2+n)(\dot{a}^{2}+a\ddot{a})^{3}(-2\dot{a}^{3}-a\dot{a}\ddot{a}
\\+a^{2}\dddot{a})^{2}+a^{2}(\lambda^{2}a^{4}+36(\dot{a}+a\ddot{a})^{2})^{2}(3\ddot{a}+4\dot{a}\ddot{a}+a\ddot{a})-72(1+n)(\dot{a}^{2}+a\ddot{a})(\lambda^{2}a^{4}+
36\dot{a}^{4}+72a\dot{a}^{2}\ddot{a}+36a^{2}\\\ddot{a}^{2})(10\dot{a}^{6}-6a\dot{a}^{4}\ddot{a}-4a^{2}\dot{a}^{3}\dddot{a}+2a^{3}\dot{a}\ddot{a}\dddot{a}+a^{2}\dot{a}^{2}(-6\ddot{a}^{2}+a\dddot{a})
+a^{3}(\ddot{a}^{3}+a\dddot{a}^{2}+a\ddot{a}\dddot{a}))))+4\Big[72n\alpha\lambda^{4}a^{8}\dot{a}^{2}\\(\dot{a}^{2}+a\ddot{a})(1+\frac{36(\dot{a}^{2}+a\ddot{a})^{2}}{\lambda^{2}a^{4}})^{-1-n}
-24n\alpha\lambda^{4}a\lambda^{9}\ddot{a}(\dot{a}^{2}+a\ddot{a})(1+\frac{36(\dot{a}^{2}+a\ddot{a})^{2}}{\lambda^{2}a^{4}})^{-1-n}+24\lambda^{4}a^{8}(\dot{a}^{2}-a\ddot{a})\\(\lambda a^{2}+12n\alpha(\dot{a}^{2}+a\ddot{a})(1+\frac{36(\dot{a}^{2}+a\ddot{a})^{2}}{\lambda^{2}a^{4}})^{-1-n})\lambda^{4}\gamma a^{8}(6\lambda a^{2}(\dot{a}^{2}+a\ddot{a})^{2}-6\dot{a}(\lambda a^{2}+12n\alpha(\dot{a}^{2}+a\ddot{a})\\(1+\frac{36(\dot{a}^{2}+a\ddot{a})^{2}}{\lambda^{2}a^{4}})^{-1-n})+\alpha\lambda^{2} a^{4}(1-(1+\frac{36(\dot{a}^{2}+a\ddot{a})^{2}}{\lambda^{2}a^{4}})^{-n})-48n\alpha\lambda^{4}a^{9}\dot{a}(\dot{a}^{2}+a\ddot{a})^{2}
(1+\frac{36(\dot{a}^{2}+a\ddot{a})^{2}}{\lambda^{2}a^{4}})^{-1-n})\\(3\dot{a}\ddot{a}+a\dddot{a})(2\dot{a}^(3)-a\dot{a}\ddot{a}-a^{2}\dddot{a})+
1728n(1+n)\alpha\lambda^{2}a^{5}(\dot{a}^{2}+a\ddot{a})^{2}(1+\frac{36(\dot{a}^{2}+a\ddot{a})^{2}}{\lambda^{2}a^{4}})^{-2-n}(3\dot{a}\ddot{a}+a\dddot{a})
\\(2\dot{a}^{3}-a\dot{a}\ddot{a}-a^{2}\dddot{a})-(12n\alpha\lambda^{6}a^{11}\dot{a}(3(1+\gamma)a-2(1+3\gamma)\dot{a})(-\lambda a^{4}+36(1+2n)\dot{a}^{4}+72(1+2n)+2a\dot{a}^{2}\ddot{a}+\\36(1+2n)a^{2}\ddot{a}^{2})(1+\frac{36(\dot{a}^{2}+a\ddot{a})^{2}}{\lambda^{2}a^{4}})^{-n}
(-2\dot{a}^{3}+a\dot{a}\ddot{a}+a^{2}\dddot{a}))/(\lambda^{2}a^{4}+36\dot{a}^{4}+72a\dot{a}^{2}\ddot{a}+36a^{2}\ddot{a}^{2})^{2}+62208n\\(1+n)(2+n)\alpha(\dot{a}^{2}+a\ddot{a})^{3}(1+\frac{36(\dot{a}^{2}+a\ddot{a})^{2}}{\lambda^{2}a^{4}})(-2\dot{a}^{3}+a\dot{a}\ddot{a}-a^{2}\dddot{a})+12n\alpha\lambda^{4}a^{10}
(1+\frac{36(\dot{a}^{2}+a\ddot{a})^{2}}{\lambda^{2}a^{4}})^{-1-n}\\(3\ddot{a}^{2}+4\dot{a}\ddot{a}+a\dddot{a})-846n(1+n)\alpha\lambda^{2}a^{4}
(\dot{a}^{2}+a\ddot{a})(1+\frac{36(\dot{a}^{2}+a\ddot{a})^{2}}{\lambda^{2}a^{4}})^{-2-n}(10\dot{a}^{6}-6a\dot{a}^{4}\ddot{a}-4a^{2}\dot{a}^{3}
+2a^{3}\dot{a}\ddot{a}\dddot{a}\\+a\dddot{a}+a^{2}\dot{a}^{2}(-6\ddot{a}^{2}+a\ddddot{a})+a^{3}(\ddot{a}^{3}+a\dddot{a}+a\dot{a}\ddot{a}))+36n\alpha\gamma
(1+\frac{36(\dot{a}^{2}+a\ddot{a})^{2}}{\lambda^{2}a^{4}}^{-3-n})\Big(6\dot{a}^{2}(36(\dot{a}^{2}+a\ddot{a})^{2})\\(\lambda^{2}a^{4}+36(\dot{a}^{2}+a\ddot{a})^{2})^{2}
-a\dot{a}(\lambda^{2}a^{4}+36(\dot{a}^{2}+a\ddot{a})^{2})^{2}(3\dot{a}\ddot{a}+a\ddot{a})-288(1+n)\dot{a}(\dot{a}^{2}+a\ddot{a})^{2}[\lambda^{2}a^{4}+
(36(\dot{a}^{2}\\+a\ddot{a})^{2})(2\dot{a}^{3}+a\dot{a}\ddot{a}-a^{2}\dddot{a})+144(1+n)a(\dot{a}+a\ddot{a})(\lambda^{2}a^{4}+36(\dot{a}^{2}+a\ddot{a})^{2})
(3\dot{a}\ddot{a}+a\dddot{a})(2\dot{a}^{3}+a\dot{a}\ddot{a}-a^{2}\dddot{a})\\+5184(1+n)(2+n)(\dot{a}^{2}+a\ddot{a})^{3}(-2\dot{a}^{3}+a\dot{a}\ddot{a}+a^{2}\dddot{a})
+[10\dot{a}^{6}-6a\dot{a}^{4}\ddot{a}-4a^{2}\dot{a}^{3}\dddot{a}+2a^{2}\dot{a}\ddot{a}\dddot{a}+a^{3}(\ddot{a}\\+a\dddot{a}^{2}+a\ddot{a}\ddddot{a})]]\Big)\Big]=0.
\end{split}
\end{equation}
\\
\\
\begin{equation}\label{16a}
\begin{split}
\frac{1}{3\lambda^{5}(1+\gamma)[3+\gamma(13+8\gamma)]a^{12}}(-\lambda^{4}a^{8}(2(2(1-\gamma)\gamma\dot{a}^{2}-(9+\gamma(19+8\gamma))a\ddot{a})(\lambda a^{2}+12n\alpha(\dot{a}^{2}+a\ddot{a})(1+\\\frac{36(\dot{a}^{2}+a\ddot{a})^{2}}{\lambda^{2}a^{4}})^{-1-n}\lambda(3+\gamma)(1+2\gamma)a^{2}
(6(\dot{a}^{2}+a\ddot{a})+\alpha\lambda a^{2}(1-(1+\frac{36(\dot{a}^{2}+a\ddot{a})^{2}}{\lambda^{2}a^{4}})^{-n})-(24n\alpha\lambda^{2}a^{3}\dot{a}(9(1+\gamma)^{2}\\a-2(-1+\gamma)\gamma\dot{a})
(-\lambda^{2}a^{4}+36(1+2n)a^{4}+72(1+2n)aa^{2}\ddot{a}^{2})(1+\frac{36(\dot{a}^{2}+a\ddot{a})^{2}}{\lambda^{2}a^{4}})^{-n}(-2\dot{a}+a\dot{a}\ddot{a}+a^{2}\dddot{a})/\lambda^{2}a^{4}\\
+36\dot{a}^{4}+72a\dot{a}\ddot{a}+36a^{2}\ddot{a}^{2})^{2})+24n\alpha(1-\gamma)\gamma(1+\frac{36(\dot{a}^{2}+a\ddot{a})^{2}}{\lambda^{2}a^{4}})^{-3-n}(6\dot{a}^{2}(\dot{a}^{2}+a\ddot{a})(\lambda^{2}a^{4}
+36(\dot{a}^{2}+a\ddot{a})^{2})^{2}\\-2a\ddot{a}(\dot{a}^{2}+a\ddot{a})(\lambda^{2}a^{4}+36(\dot{a}^{2}+a\ddot{a})^{2})^{2}-4a\dot{a}(\lambda^{2}a^{4}+36(\dot{a}^{2}+a\ddot{a})^{2})^{2}
(3\dot{a}\ddot{a}+a\dddot{a})-288(1+n)\dot{a}(\dot{a}^{2}+a\ddot{a})^{2}(\lambda^{2}a^{4}+\\36(\dot{a}^{2}+a\ddot{a})^{2})(2\dot{a}^{3}-a\dot{a}\ddot{a}-a^{2}\dddot{a})+144(1+n)a(\dot{a}^{2}+a\ddot{a})
(\lambda^{2}a^{4}+36(\dot{a}^{2}+a\ddot{a})^{2})+(3\dot{a}\ddot{a}+a\dddot{a})(2\dot{a}^{3}-a\dot{a}\ddot{a}-a^{2}\dddot{a})\\+5184(1+n)(2+n)(\dot{a}^{2}+a\ddot{a})^{3}(-2\dot{a}^{3}+a\dot{a}\ddot{a}+a^{2}
\dddot{a})^{2}+a^{2}(\lambda^{2}a^{4}+36(\dot{a}^{2}+a\ddot{a})^{2})^{2}(3\ddot{a}^{2}+4\dot{a}\ddot{a}+a\dddot{a})-72(1+n)\\(\dot{a}^{2}+a\ddot{a})\lambda^{2}a^{4}+36\dot{a}^{4}+72a\dot{a}^{2}\ddot{a}+36a^{2}\ddot{a}^{2}
(10\dot{a}^{6}-6a\dot{a}^{4}\ddot{a}-4a^{2}\dot{a}^{3}\dddot{a}+2a^{3}\dot{a}\ddot{a}\dddot{a}+a^{2}\dot{a}^{2}(-6\ddot{a}^{2}+a\dddot{a})+a^{3}(\ddot{a}^{3}+a\dddot{a}\\+a\ddot{a}\dddot{a})))+
6\gamma(72n\alpha\lambda^{4}a^{8}\dot{a}^{2}(\dot{a}^{2}+a\ddot{a})(1+\frac{36(\dot{a}^{2}+a\ddot{a})^{2}}{\lambda^{2}a^{4}})^{-1-n}+2\lambda^{4}a^{8}(\dot{a}^{2}-a\ddot{a})\lambda a^{2}+12n\alpha(\dot{a}^{2}+a\ddot{a})(1+\\\frac{36(\dot{a}^{2}+a\ddot{a})^{2}}{\lambda^{2}a^{4}})^{-1-n}-\lambda^{4}\gamma a^{8}(6\lambda a^{2}(\dot{a}^{2}+a\ddot{a})-6\dot{a}^{2}(\lambda a^{2}+12n\alpha(\dot{a}^{2}+a\ddot{a})(1+\frac{36(\dot{a}^{2}+a\ddot{a})^{2}}{\lambda^{2}a^{4}})^{-1-n}+\alpha\lambda^{2}a^{4}(1-(1+\\\frac{36(\dot{a}^{2}+a\ddot{a})^{2}}{\lambda^{2}a^{4}})^{-n}))-48n\alpha\lambda^{4}
a^{9}\dot{a}(1+\frac{36(\dot{a}^{2}+a\ddot{a})^{2}}{\lambda^{2}a^{4}})^{-1-n}(36\dot{a}\ddot{a}+a\dddot{a})-3456n(1+n)a\lambda^{2} a^{4}\dot{a}(\dot{a}^{2}+a\ddot{a})^{2}(1+\\\frac{36(\dot{a}^{2}+a\ddot{a})^{2}}{\lambda^{2}a^{4}})^{2}(2\dot{a}^{3}-a\dot{a}\ddot{a}-a^{2}\dddot{a})+1728n(1+n)
a\lambda^{2}a^{5}(\dot{a}^{2}+a\ddot{a})
(1+\frac{36(\dot{a}^{2}+a\ddot{a})^{2}}{\lambda^{2}a^{4}})^{-2-n}(3\dot{a}\ddot{a}+a\dddot{a})(2\dot{a}^{3}-a\dot{a}\ddot{a}\\-a^{2}\dddot{a})-(12n\alpha\lambda
^{6}a^{11}\dot{a}\Big(3(1+\gamma)a-2(1+3\gamma)\dot{a}\Big[-\lambda^{2}a^{4}+36(1+2n)\dot{a}^{4}
+72(1+2n)\dot{a}^{2}\ddot{a}+36(1+2n)a^{2}\dot{a}^{2}(1+\\\frac{36(\dot{a}^{2}+a\ddot{a})^{2}}{\lambda^{2}a^{4}})^{-n}(-2\dot{a}^{3}+a\dot{a}\ddot{a}+a^{2}\dddot{a})
/(\lambda^{2}a^{4}+36\dot{a}^{4}+72a\dot{a}^{2}\ddot{a}
+36a^{2}\ddot{a})^{2}+62208n(1+n)(2+n)\alpha(\dot{a}^{2}+a\ddot{a})^{3}(1+\\\frac{36(\dot{a}^{2}+a\ddot{a})^{2}}{\lambda^{2}a^{4}})^{-3-n}(-2\dot{a}^{3}+a\dot{a}
\ddot{a}+a^{2}\dddot{a})+12n\alpha\lambda^{4}a^{10}
(1+\frac{36(\dot{a}^{2}+a\ddot{a})^{2}}{\lambda^{2}a^{4}})^{1-n}(3\ddot{a}^{2}+4\dot{a}\dddot{a}+a\ddddot{a})-864n(1+n)\alpha\\\lambda^{2}a^{4}(\dot{a}^{2}
+a\ddot{a})(1+\frac{36(\dot{a}^{2}+a\ddot{a})^{2}}{\lambda^{2}a^{4}})^{-2-n}
(10\dot{a}^{6}-6a\dot{a}^{4}\ddot{a}-4a^{2}dot{a}^{3}\dddot{a}+2a^{2}\dot{a}\ddot{a}\dddot{a}+a^{2}\dot{a}^{2}(-6\ddot{a}+a\ddddot{a})+a^{3}(\ddot{a}^{3}+
a\dddot{a}^{2}\\+a\ddot{a}\ddddot{a}))+36n\alpha\gamma
(1+\frac{36(\dot{a}^{2}+a\ddot{a})^{2}}{\lambda^{2}a^{4}})^{-3-n}(6\dot{a}^{2}(\dot{a}^{2}+a\ddot{a})(\lambda^{2}a^{4}+36(\dot{a}^{2}+a\ddot{a})^{2})^{2}
-2a\ddot{a}(\dot{a}^{2}+a\ddot{a})(\lambda^{2}a^{4}+36(\dot{a}^{2}+\\a\ddot{a})^{2})^{2}
-4a\dot{a}(\lambda^{2}a^{4}+36(\dot{a}^{2}+a\ddot{a})^{2})^{2}(3\dot{a}\ddot{a}+a\dddot{a})-288(1+n)\dot{a}(\dot{a}^{2}+a\ddot{a})^{2}(\lambda^{2}a^{4}+36(
\dot{a}^{2}+a\ddot{a})^{2})(2\dot{a}^{3}-a\dot{a}\ddot{a}-a^{2}\dddot{a})\\+144(1+n)
a(\dot{a}^{2}+a\ddot{a})(\lambda^{2}a^{4}+36(\dot{a}^{2}+a\ddot{a})^{2})(3\dot{a}\ddot{a}+a\dddot{a})(2\dot{a}^{3}-a\dot{a}\ddot{a}-a^{2}\dddot{a})+5184(1+n)(2+n)
(\dot{a}^{2}+a\ddot{a})^{2})\\(3\dot{a}\ddot{a}+a\dddot{a})(2\dot{a}^{3}-a\dot{a}\ddot{a}-a^{2}\dddot{a})^{2}(3\ddot{a}^{2}+4\dot{a}\dddot{a}+a\ddddot{a})-72(1+n)
(\dot{a}^{2}+a\ddot{a})(\lambda^{2}a^{4}+36\dot{a}^{4}
+72\dot{a}^{2}\ddot{a}+36a^{2}\ddot{a}^{2})[10\dot{a}^{6}\\-6a\dot{a}^{4}\ddot{a}-4a^{2}\dot{a}\dddot{a}+2a^{3}\dot{a}\ddot{a}\dddot{a}+a^{2}\dot{a}^{2}
(-6\ddot{a}^{2}+a\ddddot{a})+a^{3}(\ddot{a}^{3}+a\dddot{a}^{2}+a\ddot{a}\dddot{a})]\Big]\Big)=0.
\end{split}
\end{equation}
\\
\\
\begin{equation}\label{30a}
\begin{split}
\frac{1}{4\lambda^{5}(1+\gamma)[3+\gamma(13+8\gamma)]a^{12}}(-\lambda^{4}a^{8}(2(2(1-\gamma)\gamma\dot{a}^{2}-(9+\gamma(19+8\gamma))a\ddot{a})(\lambda a^{2}+12n\alpha(\dot{a}^{2}+a\ddot{a})(1+\\\frac{36(\dot{a}^{2}+a\ddot{a})^{2}}{\lambda^{2}a^{4}})^{-1-n}\lambda(3+\gamma)(1+2\gamma)a^{2}
(6(\dot{a}^{2}+a\ddot{a})+\alpha\lambda a^{2}(1-(1+\frac{36(\dot{a}^{2}+a\ddot{a})^{2}}{\lambda^{2}a^{4}})^{-n})-(24n\alpha\lambda^{2}a^{3}\dot{a}(9(1+\gamma)^{2}\\a-2(-1+\gamma)\gamma\dot{a})
(-\lambda^{2}a^{4}+36(1+2n)a^{4}+72(1+2n)aa^{2}\ddot{a}^{2})(1+\frac{36(\dot{a}^{2}+a\ddot{a})^{2}}{\lambda^{2}a^{4}})^{-n}(-2\dot{a}+a\dot{a}\ddot{a}+a^{2}
\dddot{a})/\lambda^{2}a^{4}\\
+36\dot{a}^{4}+72a\dot{a}\ddot{a}+36a^{2}\ddot{a}^{2})^{2})+24n\alpha(1-\gamma)\gamma(1+\frac{36(\dot{a}^{2}+a\ddot{a})^{2}}{\lambda^{2}a^{4}})^{-3-n}(6\dot{a}^{2}
(\dot{a}^{2}+a\ddot{a})(\lambda^{2}a^{4}
+36(\dot{a}^{2}+a\ddot{a})^{2})^{2}\\-2a\ddot{a}(\dot{a}^{2}+a\ddot{a})(\lambda^{2}a^{4}+36(\dot{a}^{2}+a\ddot{a})^{2})^{2}-4a\dot{a}(\lambda^{2}a^{4}+36
(\dot{a}^{2}+a\ddot{a})^{2})^{2}
(3\dot{a}\ddot{a}+a\dddot{a})-288(1+n)\dot{a}(\dot{a}^{2}+a\ddot{a})^{2}(\lambda^{2}a^{4}+\\36(\dot{a}^{2}+a\ddot{a})^{2})(2\dot{a}^{3}-a\dot{a}\ddot{a}-a^{2}
\dddot{a})+144(1+n)a(\dot{a}^{2}+a\ddot{a})
(\lambda^{2}a^{4}+36(\dot{a}^{2}+a\ddot{a})^{2})+(3\dot{a}\ddot{a}+a\dddot{a})(2\dot{a}^{3}-a\dot{a}\ddot{a}-a^{2}\dddot{a})\\+5184(1+n)(2+n)(\dot{a}^{2}
+a\ddot{a})^{3}(-2\dot{a}^{3}+a\dot{a}\ddot{a}+a^{2}
\dddot{a})^{2}+a^{2}(\lambda^{2}a^{4}+36(\dot{a}^{2}+a\ddot{a})^{2})^{2}(3\ddot{a}^{2}+4\dot{a}\ddot{a}+a\dddot{a})-72(1+n)\\(\dot{a}^{2}+a\ddot{a})\lambda^{2}
a^{4}+36\dot{a}^{4}+72a\dot{a}^{2}\ddot{a}+36a^{2}\ddot{a}^{2}
(10\dot{a}^{6}-6a\dot{a}^{4}\ddot{a}-4a^{2}\dot{a}^{3}\dddot{a}+2a^{3}\dot{a}\ddot{a}\dddot{a}+a^{2}\dot{a}^{2}(-6\ddot{a}^{2}+a\dddot{a})+a^{3}(\ddot{a}^{3}
+a\dddot{a}\\+a\ddot{a}\dddot{a})))+
6\gamma(72n\alpha\lambda^{4}a^{8}\dot{a}^{2}(\dot{a}^{2}+a\ddot{a})(1+\frac{36(\dot{a}^{2}+a\ddot{a})^{2}}{\lambda^{2}a^{4}})^{-1-n}+2\lambda^{4}a^{8}(\dot{a}^{2}
-a\ddot{a})\lambda a^{2}+12n\alpha(\dot{a}^{2}+a\ddot{a})(1+\\\frac{36(\dot{a}^{2}+a\ddot{a})^{2}}{\lambda^{2}a^{4}})^{-1-n}-\lambda^{4}\gamma a^{8}(6\lambda a^{2}(\dot{a}^{2}+a\ddot{a})-6\dot{a}^{2}(\lambda a^{2}+12n\alpha(\dot{a}^{2}+a\ddot{a})(1+\frac{36(\dot{a}^{2}+a\ddot{a})^{2}}{\lambda^{2}a^{4}})^{-1-n}+\alpha\lambda^{2}a^{4}(1-(1+\\\frac{36(\dot{a}^{2}+a\ddot{a})^{2}}{\lambda^{2}a^{4}})^{-n}))-48n\alpha\lambda^{4}
a^{9}\dot{a}(1+\frac{36(\dot{a}^{2}+a\ddot{a})^{2}}{\lambda^{2}a^{4}})^{-1-n}(36\dot{a}\ddot{a}+a\dddot{a})-3456n(1+n)a\lambda^{2} a^{4}\dot{a}(\dot{a}^{2}+a\ddot{a})^{2}(1+\\\frac{36(\dot{a}^{2}+a\ddot{a})^{2}}{\lambda^{2}a^{4}})^{2}(2\dot{a}^{3}-a\dot{a}\ddot{a}-a^{2}\dddot{a})+1728n(1+n)
a\lambda^{2}a^{5}(\dot{a}^{2}+a\ddot{a})
(1+\frac{36(\dot{a}^{2}+a\ddot{a})^{2}}{\lambda^{2}a^{4}})^{-2-n}(3\dot{a}\ddot{a}+a\dddot{a})(2\dot{a}^{3}-a\dot{a}\ddot{a}\\-a^{2}\dddot{a})-(12n\alpha\lambda^{6}
a^{11}\dot{a}\Big(3(1+\gamma)a-2(1+3\gamma)\dot{a}[-\lambda^{2}a^{4}+36(1+2n)\dot{a}^{4}
+72(1+2n)\dot{a}^{2}\ddot{a}+36(1+2n)a^{2}\dot{a}^{2}(1+\\\frac{36(\dot{a}^{2}+a\ddot{a})^{2}}{\lambda^{2}a^{4}})^{-n}(-2\dot{a}^{3}+a\dot{a}\ddot{a}+a^{2}
\dddot{a})/(\lambda^{2}a^{4}+36\dot{a}^{4}+72a\dot{a}^{2}\ddot{a}
+36a^{2}\ddot{a})^{2}+62208n(1+n)(2+n)\alpha(\dot{a}^{2}+a\ddot{a})^{3}(1+\\\frac{36(\dot{a}^{2}+a\ddot{a})^{2}}{\lambda^{2}a^{4}})^{-3-n}(-2\dot{a}^{3}
+a\dot{a}\ddot{a}+a^{2}\dddot{a})+12n\alpha\lambda^{4}a^{10}
(1+\frac{36(\dot{a}^{2}+a\ddot{a})^{2}}{\lambda^{2}a^{4}})^{1-n}(3\ddot{a}^{2}+4\dot{a}\dddot{a}+a\ddddot{a})-864n(1+n)\alpha\\\lambda^{2}a^{4}(\dot{a}^{2}
+a\ddot{a})(1+\frac{36(\dot{a}^{2}+a\ddot{a})^{2}}{\lambda^{2}a^{4}})^{-2-n}
(10\dot{a}^{6}-6a\dot{a}^{4}\ddot{a}-4a^{2}dot{a}^{3}\dddot{a}+2a^{2}\dot{a}\ddot{a}\dddot{a}+a^{2}\dot{a}^{2}(-6\ddot{a}+a\ddddot{a})+a^{3}(\ddot{a}^{3}
+a\dddot{a}^{2}\\+a\ddot{a}\ddddot{a}))+36n\alpha\gamma
(1+\frac{36(\dot{a}^{2}+a\ddot{a})^{2}}{\lambda^{2}a^{4}})^{-3-n}(6\dot{a}^{2}(\dot{a}^{2}+a\ddot{a})(\lambda^{2}a^{4}+36(\dot{a}^{2}+a\ddot{a})^{2})^{2}-
2a\ddot{a}(\dot{a}^{2}+a\ddot{a})(\lambda^{2}a^{4}+36(\dot{a}^{2}+\\a\ddot{a})^{2})^{2}
-4a\dot{a}(\lambda^{2}a^{4}+36(\dot{a}^{2}+a\ddot{a})^{2})^{2}(3\dot{a}\ddot{a}+a\dddot{a})-288(1+n)\dot{a}(\dot{a}^{2}+a\ddot{a})^{2}(\lambda^{2}a^{4}
+36(\dot{a}^{2}+a\ddot{a})^{2})(2\dot{a}^{3}-a\dot{a}\ddot{a}-a^{2}\dddot{a})\\+144(1+n)
a(\dot{a}^{2}+a\ddot{a})(\lambda^{2}a^{4}+36(\dot{a}^{2}+a\ddot{a})^{2})(3\dot{a}\ddot{a}+a\dddot{a})(2\dot{a}^{3}-a\dot{a}\ddot{a}-a^{2}\dddot{a})+5184(1+n)(2+n)
(\dot{a}^{2}+a\ddot{a})^{2})\\(3\dot{a}\ddot{a}+a\dddot{a})(2\dot{a}^{3}-a\dot{a}\ddot{a}-a^{2}\dddot{a})^{2}(3\ddot{a}^{2}+4\dot{a}\dddot{a}+a\ddddot{a})
-72(1+n)(\dot{a}^{2}+a\ddot{a})(\lambda^{2}a^{4}+36\dot{a}^{4}
+72\dot{a}^{2}\ddot{a}+36a^{2}\ddot{a}^{2})[10\dot{a}^{6}\\-6a\dot{a}^{4}\ddot{a}-4a^{2}\dot{a}\dddot{a}+2a^{3}\dot{a}\ddot{a}\dddot{a}+a^{2}\dot{a}^{2}
(-6\ddot{a}^{2}+a\ddddot{a})+a^{3}(\ddot{a}^{3}+a\dddot{a}^{2}+a\ddot{a}\dddot{a})]\}
\Big)=0.
\end{split}
\end{equation}
\\
\\
\begin{equation}\label{31a}
\begin{split}
\frac{-1}{3+\gamma(13+8\gamma)}[3\alpha\lambda+\frac{18(\dot{a}^2+a\ddot{a})}{a^{2}}
-\alpha\lambda(3+4\gamma)(1+\frac{36(\dot{a}^2+a\ddot{a})^{2}}{\lambda^{2}a^{4}})^{-n}-\frac{1}{\lambda a^{3}} (18\ddot{a}(\lambda a^{2}
+12n\alpha(\dot{a}^2+a\ddot{a})\\(1+\frac{36(\dot{a}^2+a\ddot{a})^{2}}{\lambda^{2} a^{4}})^{-1-n}))-
8\gamma(-\dot{a}^2-2a\ddot{a})(\lambda\alpha^{2}+12n\alpha(\dot{a}^2+a\ddot{a})(1+\frac{36(\dot{a}^2+a\ddot{a})^{2}}{\lambda^{2}a^{4}})^{-1-n})-
(24n\alpha\lambda\dot{a}((9+6\gamma)a\\+4\gamma\dot{a})
(-\lambda^{2} a^{4}+36(1+2n)\dot{a}^{4}++72(1+2n)a\dot{a}^{2}\ddot{a}+36(1+2n)a^{2}\ddot{a}^{2})(1+\frac{36(\dot{a}^2+a\ddot{a})^{2}}{\lambda^{2}
a^{4}})^{-n}(-2\dot{a}^{3}+a\dot{a}\ddot{a}+a^{2}\dddot{a}))/\\(a(\lambda^{2}a^{4}+36\dot{a}^{4}+72a\dot{a}^{2}\ddot{a}+36a^{2}\ddot{a}^{2})^{2})
+\frac{1}{\lambda^{5} a^{12}}(\alpha\lambda^{6}a^{12}+6\lambda^{5}a^{10}(\dot{a}^2+a\ddot{a})-72\alpha\lambda^{4}a^{8}\dot{a}(\dot{a}^2+a\ddot{a})a(1+\\\frac{36(\dot{a}^2+a\ddot{a})^{2}}
{\lambda^{2}
a^{4}})^{-1-n}+24n\alpha\lambda^{4}a^{9}\ddot{a}(\dot{a}^2+a\ddot{a})(1+\frac{36(\dot{a}^2+a\ddot{a})^{2}}{\lambda^{2}
a^{4}})^{-1-n}+48n\alpha\lambda^{4}a^{9}\dot{a}(1+\frac{36(\dot{a}^2+a\ddot{a})^{2}}{\lambda^{2}
a^{4}})^{-1-n}(3\dot{a}^{3}\ddot{a}\\+a\dddot{a})+3456n(1+n)\alpha\lambda^{2}a^{4}\dot{a}(\dot{a}^2+a\ddot{a})(1+\frac{36(\dot{a}^2+a\ddot{a})^{2}}{\lambda^{2}
a^{4}})^{-2-n}(2\dot{a}^{3}-a\dot{a}\ddot{a}-a^{2}\dddot{a})-\Big(2\dot{a}^{3}-1728n(1+n)\alpha\lambda^{2}a^{5}\\(\dot{a}^2+a\ddot{a})(1+\frac{36(\dot{a}^2+a\ddot{a}
)^{2}}{\lambda^{2}
a^{4}})^{-2-n}(3\dot{a}^{3}\ddot{a}+a\dddot{a})(2\dot{a}^{3}-a\dot{a}\ddot{a}-a^{2}\dddot{a})-62280n(1+n)(2+n)\alpha[\dot{a}^2+a\ddot{a}(1+\\\frac{36(\dot{a}^2+
a\ddot{a})^{2}}{\lambda^{2}
a^{4}})^{-3-n}(-2\dot{a}^{3}+a\dot{a}\ddot{a}+a^{2}\dddot{a})-12n\alpha\lambda^{4}a^{10}(1+\frac{36(\dot{a}^2+a\ddot{a})^{2}}{\lambda^{2}
a^{4}})^{-2-n}(3\ddot{a}^{2}+4\dot{a}\ddot{a}+a\dddot{a})+864n(1+n)\alpha\\ \lambda^{2}a^{4}(\dot{a}^2+a\ddot{a})(1+\frac{36(\dot{a}^2+a\ddot{a})^{2}}{\lambda^{2}
a^{4}})^{-2-n}
[10\dot{a}^{6}-6a\dot{a}^{4}\ddot{a}-4a^{2}\dot{a}^{3}\dddot{a}+2a^{3}\dot{a}\ddot{a}\dddot{a}+a^{2}\dot{a}^{2}(-6\dot{a}^{2}+a\dddot{a})+a^{3}
(\ddot{a}+a\dddot{a}^{2}+\\a\ddot{a}\dddot{a})]]\Big)=0,
\end{split}
\end{equation}

%

\section*{Contributions}
\hskip\parindent
\small
 We declare that all the authors have same contributions to this paper.

\section*{Data Availability Statement}
The authors declare that the data supporting the findings of this study are available within the article.

\section*{Acknowledgement}
Adnan Malik acknowledges the Grant No. YS304023912 to support his Postdoctoral Fellowship at Zhejiang Normal University, China.

\title
\end{document}